\documentclass[10pt,twocolumn]{article}

\usepackage[boxruled,noend,vlined]{algorithm2e}
\SetKwFor{ParallelForEach}{parallel foreach}{do}{}
\SetAlgoSkip{} 
\newcommand{\Set}{\mathrel{\mathop:}=}

\usepackage{url}

\usepackage[colorlinks,bookmarks=false]{hyperref}

\begin{document}

\title{Faster Radix Sort via Virtual Memory and Write-Combining}
\author{
Jan Wassenberg\\Fraunhofer IOSB\\{\tt\small {jan.wassenberg@iosb.fraunhofer.de}}
\and Peter Sanders\\KIT
}
\maketitle

\pdfinfo{
   /Author (Jan Wassenberg, Peter Sanders)
   /Title (Faster Radix Sort via Virtual Memory and Write-Combining)
   /CreationDate (D:20100817102800)
   /Subject ()
   /Keywords ()
   /Creator ()
   /Producer ()
}

\begin{abstract}

Sorting algorithms are the deciding factor for the performance of
common operations such as removal of duplicates or database sort-merge joins.
This work focuses on 32-bit integer keys, optionally paired with a
32-bit value. We present a fast radix sorting algorithm that builds
upon a microarchitecture-aware variant of counting sort.
Taking advantage of virtual memory and making use of write-combining
yields a per-pass throughput corresponding to at least 88~\% of the
system's peak memory bandwidth. Our implementation outperforms Intel's
recently published radix sort by a factor of 1.5. It also compares
favorably to the reported performance of an algorithm for Fermi~GPUs
when data-transfer overhead is included. These results indicate that
scalar, bandwidth-sensitive sorting algorithms remain competitive on
current architectures. Various other memory-intensive applications
can benefit from the techniques described herein.
\end{abstract}

\section{Introduction}
\label{sec:intro}

Sorting is a fundamental operation that is a time-critical component
of various applications such as databases and search engines.
The well-known lower bound of $\Omega(n \cdot \log n)$ for comparison-based
algorithms no longer applies when special properties of the keys
can be assumed. In this work, we focus on 32-bit integer keys,
optionally paired with a 32-bit value (though larger sizes are possible).
This simplifies the implementation without loss of generality, since
applications can often replace large records with a pointer or
index \cite{partialKey}.
The radix sort algorithm is commonly used in such cases due to its
$O(N)$ complexity. In this preliminary report, we show a 1.5-fold
performance increase over results recently published by Intel \cite{intelSort}.

The remaining sections are organized in a bottom-up fashion,
with Section~\ref{sec:swwc} dedicated to the basic realities of
current and future microarchitectures that affect memory-intensive
programs and motivate our approach.
We build upon this foundation in Section~\ref{sec:counting},
showing how to speed up counting sort by taking advantage of
virtual memory and write-combining. Section~\ref{sec:radix}
applies this technique towards our main contribution, a novel
variant of radix sort. The performance of our implementation
is evaluated in Section~\ref{sec:perf}.
Bandwidth measurements indicate the per-pass throughput is
nearly optimal for the given hardware. Its two CPUs outperform
a Fermi GPU when accounting for data-transfer overhead.

\section{Software Write-Combining}
\label{sec:swwc}

We begin with a description of basic microarchitectural realities
that are likely to have a serious impact on applications with
numerous memory accesses, and show how to avoid performance penalties
by means of Software Write-Combining. These topics are not new, but
we believe they are often not adequately addressed.

The first problem arises when writing items to multiple streams.
An ideal cache with at least as many lines could exploit the writes'
spatial locality and entirely avoid noncompulsory misses. However,
perfect hit rates are not achievable in practice due to limited
ways of associativity $a$ \cite{sandersB1a}. Since only $a$ lines
can be mapped to a cache set, any further allocations from that
set result in the eviction of one of the previous lines. If possible,
applications should avoid writing to many different streams. Otherwise,
the various write positions should map to different sets to
avoid thrashing and conflict misses. For current L1 caches with
$a=8$~ways, size $C=32$~KiB and lines of $B=64$~bytes, there are
$S = \frac{C}{a \cdot B} = 64$~sets, and bits $\left[\lg B, \lg B + \lg S\right)$
of the destination addresses should differ (e.g.\ by ensuring the
write positions are not a multiple of $S \cdot B = 4$~KiB apart).

A second issue is provoked by a large number of write-only accesses.
Even if an entire cache line is to be written, the previous destination
memory must first be read into the cache. While the corresponding latency
may be partially hidden via prefetching, the cache line allocations
remain problematic due to capacity constraints and eviction policy.
Instead of displacing write-only lines that are not accessed after
having been filled, the widespread (pseudo-)Least-Recently-Used strategy
displaces previously cached data due to their older timestamp.
An attempt to avoid these evictions by explicitly invalidating
cache lines (e.g.\ with the IA-32 CLFLUSH instruction) did not
yield meaningful improvements. Instead, applications should avoid
`cache pollution' by writing directly to memory via non-temporal
streaming stores.

This leads directly to the next concern: single memory accesses
involve significant bus overhead. The architecture therefore combines
neighboring non-temporal writes into a single burst transfer.
However, currently microarchitectures only provide four to ten
write-combine (WC) buffers \cite{intelSysProg}. Non-temporal writes to
multiple streams may force these buffers to be flushed to memory via
`partial writes' before they are full. The application can prevent this
by making use of Software Write-Combining \cite{intelOpt}. The data
to be written is first placed into temporary buffers, which almost
certainly reside in the cache because they are frequently accessed.
When full, a buffer is copied to the actual destination via consecutive
non-temporal writes, which are guaranteed to be combined into a single
burst transfer.

This scheme avoids reading the destination memory, which may incur
relatively expensive Read-For-Ownership transactions and would only
pollute the cache. It works around the limited number of WC buffers by
using L1 cache lines for that purpose. Interestingly, this is tantamount
to direct software control of the transparently managed cache.

We recommend the use of such Software Write-Combining whenever a core's
active write destinations outnumber its write-combine buffers.
Fortunately, this can be done at a fairly high level, since only the
buffer copying requires special vector loads and non-temporal stores
(which are best expressed by the SSE2 intrinsics built into the
major compilers).

\section{Virtual-Memory Counting Sort}
\label{sec:counting}

We now review Counting Sort and describe an improved variant
that makes use of virtual memory and write-combining.

The na{\"i}ve algorithm first generates a histogram of the $N$ keys.
After computing the prefix sum to yield the starting output location
for each key, each value is written at its key's output position,
which is subsequently incremented.

Our first optimization goal is to avoid the initial counting pass.
We could instead insert each value into a per-key container,
e.g.\ a list of buckets. However, this incurs some overhead for
checking whether the current bucket is full.
A large array of $M$ pre-allocated buckets is more efficient, because items
can simply be written to the next free position (c.f.\ Algorithm~\ref{alg:counting},
introduced in \cite{phmsfFull}).
\begin{algorithm}[ht]
\SetKwData{Storage}{storage}
\SetKwData{Next}{next}
\SetKwData{Key}{key}
\SetKwData{Value}{value}
$\Storage \Set \FuncSty{ReserveAddressSpace}(N \cdot M)$\;
\lFor{$i \Set 0 \; \KwTo \; M$}{$\Next\left[i\right] \Set i \cdot N$}\;
\ForEach{\Key,\Value}
{
  $\Storage\left[\Next\left[\Key\right]\right] \Set \Value$\;
  $\Next\left[\Key\right] \Set \Next\left[\Key\right]+1$\;
}
\caption[]{Single-pass counting sort}
\label{alg:counting}
\end{algorithm}%
This algorithm only writes and reads each item once, a feat that comes at
the price of $N \cdot M$ space. While this appears problematic in the
Random-Access-Machine model, it is easily handled by 64-bit CPUs with
paged virtual memory. Physical memory is only mapped to pages when they
are first accessed,\footnote{Accesses to non-present pages result in
a page fault exception. The application receives such events via signals
(POSIX) or Vectored Exception Handling (Microsoft Windows) and reacts by
committing memory, after which the faulting instruction is repeated.}
thus reducing the actual memory requirements to $O(N + M \cdot \mathrm{pageSize})$.
The remainder of the initial allocation only occupies address space,
of which multiple terabytes are available on 64-bit systems.

Having avoided the initial counting pass, we now show how to efficiently
write values to \DataSty{storage} using the write-combining technique
described in Section~\ref{sec:swwc}. Our implementation initializes the
\DataSty{next} pointers to consecutive, naturally aligned, cache-line-sized
buffers. A buffer is full when its (post-incremented) position is evenly
divisible by its size. When that happens, an unrolled loop of
non-temporal writes copies the buffer to its key's current output
position within \DataSty{storage}. These output positions are also stored in
an array of pointers.

\section{Radix Sort}
\label{sec:radix}

After a brief review of radix sorting, we introduce a new
variant based on the virtual-memory counting sort described in
Section~\ref{sec:counting}.

A radix sort successively examines $D$-bit `digits' of the $K$-bit keys.
They are characterized by the order in which digits are processed:
starting at the Least Significant Digit (LSD), or Most Significant Digit (MSD).

An MSD radix sort partitions the items according to the current digit,
then recursively sorts the resulting buckets. While it no longer needs
to move items whose previously seen key digits are unique, this is not
especially helpful when the number of passes $K/D$ is small.
In fact, the overhead of managing numerous (nearly empty) buckets makes
MSD radix sort less suited for relatively small $N$.

By contrast, each iteration of the LSD variant partitions {\em all} items
into buckets by the current key digit. Since buckets are not recursively
split, their sizes are nearly equal (under the assumption of a
uniform key distribution) and the sort is stable (preserving the
original relative order of values with equal keys). However, this
comes at the cost of more copying.

To reduce this overhead and also parallel communication, we make use of
``reverse sorting'' \cite{navarroRev}, in which one or more
MSD passes partition the data into buckets, which are then locally
sorted via LSD. This turns out to be even more advantageous for
Non-Uniform Memory Access (NUMA) systems because each processor is
responsible for writing a contiguous range of outputs, thus ensuring
the OS allocates those pages from the processor's NUMA node \cite{Mey2007}.

\begin{algorithm}[t]
\caption{Parallel Radix Sort}
\label{alg:radix}
\SetKwData{Item}{item}
\SetKwData{BucketSizes}{bucketSizes}
\SetKwData{OutputIndices}{outputIndices}
\SetKwData{HistogramTwo}{histogram2}
\SetKwData{BucketsZero}{buckets0}
\SetKwData{BucketsOne}{buckets1}
\SetKwData{BucketsThree}{buckets3}
\SetKwData{BucketZero}{bucket0}
\SetKwData{BucketOne}{bucket1}
\SetKwData{BucketTwo}{bucket2}
\SetKwData{BucketThree}{bucket3}
\SetKwData{Output}{output}
\SetKwFunction{Barrier}{Barrier}
\SetKwFunction{PrefixSum}{PrefixSum}
\SetKwFunction{Digit}{Digit}
\ParallelForEach{\Item}
{
  $d \Set \Digit(\Item, 3)$\;
  $\BucketsThree\left[d\right] \Set \BucketsThree\left[d\right] \cup \left\{\Item\right\}$\;
}
$\Barrier$\;
\ForEach{$i \in \left[0, 2^D\right)$}
{
  $\BucketSizes\left[i\right] \Set \sum_{\mathrm{PE}} \left|\BucketsThree\left[i\right]\right|$\;
}
$\OutputIndices \Set \PrefixSum(\BucketSizes)$\;
\ParallelForEach{$\BucketThree \in \BucketsThree$}
{
  \ForEach{$\Item \in \BucketThree \,\forall\, \mathrm{PE}$}
  {
    $d \Set \Digit(\Item, 0)$\;
    $\BucketsZero\left[d\right] \Set \BucketsZero\left[d\right] \cup \left\{\Item\right\}$\;
  }
  \ForEach{$\BucketZero \in \BucketsZero$}
  {
    \ForEach{$\Item \in \BucketZero$}
    {
      $d \Set \Digit(\Item, 1)$\;
      $\BucketsOne\left[d\right] \Set \BucketsOne\left[d\right] \cup \left\{\Item\right\}$\;
      $d \Set \Digit(\Item, 2)$\;
      $\HistogramTwo\left[d\right] \Set \HistogramTwo\left[d\right] + 1$\;
    }
  }
  \ForEach{$\BucketOne \in \BucketsOne$}
  {
    \ForEach{$\Item \in \BucketOne$}
    {
      $d \Set \Digit(\Item, 2)$\;
      $i \Set \OutputIndices\left[d\right] + \HistogramTwo\left[d\right]$\;
      $\HistogramTwo\left[d\right] \Set \HistogramTwo\left[d\right] + 1$\;
      $\Output\left[i\right] \Set \Item$\;
    }
  }
}
\end{algorithm}%
Let us now examine the pseudocode of the radix sort (Algorithm~\ref{alg:radix}),
choosing $K=32$ for brevity and $D=8$ to allow extracting key digits without masking.
Each Processing Element (PE) first uses counting sort to partition
its items into local buckets by the MSD (digit = 3). Note that items consist
of a key and value, which are adjacent in memory (ideally within a
native 64-bit word, but larger combinations are possible in our
implementation via larger user-defined types). After all are finished,
the output index of the first item of a given MSD is computed via
prefix sum.
Each PE is assigned a range of MSD values, sorting the buckets from all PEs
for each value. Note that skewed MSD distributions cause load imbalance,
which can be resolved by one or more additional recursive MSD passes
(left for future work). The local sort entails $K/D - 1$ iterations in
LSD order. The first copies the other PE's buckets into local memory.
Pass $K/D-1$ also computes the histogram of the final digit. This allows
writing directly to the output positions in the final pass. Note that
three sets of buckets are required, which makes heavy use of virtual memory
($3 \cdot 2^D \cdot \left|PE\right| = 6144$ times the input size).
While 64-bit Linux grants each process 128 TiB address space, Windows limits
this to 8 TiB, which means only about 1 GiB of inputs can be sorted.
This restriction can be lifted when the key distribution is known and
each bucket does not need to pre-allocate storage for all $N$ items.

We briefly discuss additional system-specific considerations.
The radix $2^D$ was motivated by easy access to each digit, but
is also limited by the cache and TLB size. Because of the many required
TLB entries, we map the buckets with small pages, for which the
Intel i7 microarchitecture has 512 second-level TLB entries.
To increase TLB coverage, we use large pages for the inputs.
The working set consists of $2^D$ buffers, buffer pointers,
output positions, and 32-bit histogram counters. This fits in
a 32 KiB L1 data cache if the software write-combine buffers are
limited to a single 64-byte cache line. To avoid associativity and
aliasing conflicts, these arrays are contiguous in memory.
Interestingly, these optimizations do not detract from the readability
of the source code. Knowledge of the microarchitecture can also be
applied towards middle-level languages and enables principled design decisions.

\section{Performance Evaluation}
\label{sec:perf}

We characterize the performance of our sorting implementation by
its throughput, defined as $\frac{N}{t_1-t_0}$, where N = 64~Mi and
$t_0$ and $t_1$ are the earliest and latest start and finish times
reported by any thread. The test platform consists of dual W5580
CPUs (3.2~GHz, 48~GiB DDR3-1066 memory) running Windows XP x64. Our
implementation is compiled with ICC~11.1.082 {\small \texttt{/Ox /Og
/Oi /Ot /Qipo /GA /EHsc /MD /GS- /fp:fast=2 /GR- /Qopenmp
/QaxSSE4.2 /Quse-intel-optimized-headers}}.
For uniformly distributed 32-bit keys generated by the WELL512 algorithm \cite{wellRNG}
and no associated values, the basic algorithm (`VM only')
reaches a throughput of 334~M/s, as shown in the second column of Table~\ref{tab:throughput}.
When write-combining is enabled (`VM+WC'), performance nearly doubles
to 621~M/s.

Intel has reported 240~M/s for the same task and a single but identical
CPU \cite{intelSort}.
For a fair comparison with our dual-CPU system, we double the given
throughput, which assumes their algorithm is NUMA-aware, 
scales perfectly and is not running at a lower memory clock (since
DDR3-1066 is at the lower end of currently available frequencies).
We must also divide by the given speedup of 1.2 due to hyperthreads,
since those are disabled on our machine. This (`Intel x2') yields 400~M/s;
the proposed algorithm is therefore more than 1.5 times as fast.

A separate publication has also presented results \cite{intelMIC} for
the Many Integrated Cores architecture. The Knights Ferry processor
provides 32 cores, each with 4 threads and 16-wide SIMD. The simulation
(`KNF MIC') shows a throughput of 560 M/s. Our scalar implementation is
currently 1.1 times as fast when running on 8 cores.

Recently, a throughput of 1005 M/s was reported on a GTX 480 (Fermi) GPU \cite{gpuSort}.
However, this excludes driver and data-transfer overhead. For applications
in which the data is generated and consumed by the CPU, we must include
at least the time required to read and write data over the PCIe 2.0 bus.
Assuming the peak per-direction bandwidth of 8 GB/s is reached, the aggregate
throughput (`GPU+PCIe') is 501 M/s. Our implementation, running on two CPUs,
therefore outperforms this algorithm on the current top-of-the-line GPU by a
factor of 1.24 despite lower transistor counts ($2\cdot 731$~M vs.\ $3000$~M)
and thermal design power ($2\cdot 130$~W vs.\ $275 .. 300$~W).

\begin{table}[htbp]
\centering
\begin{tabular}{lcc}
\hline\noalign{\smallskip}
Algorithm & ~~K=32,V=0~~ & K=32, V=32\\
\noalign{\smallskip}
\hline
\noalign{\smallskip}
VM only     & 334 & 203\\
Intel x2    & 400 & 307\\
GPU+PCIe    & 501 & 303\\
KNF MIC     & 560 & (unknown)\\
VM+WC       & 621 & 430\\
\hline
\end{tabular}%
\hfill%
\caption{Throughputs [million items per second] for
32-bit keys and optional 32-bit values.}
\label{tab:throughput}
\end{table}%

Similar measurements and extrapolations for the case of 32-bit keys associated
with $V=32$-bit values are given in the third column of Table~\ref{tab:throughput}.
Since the slowdown is less than a factor of two, the implementations are at
least partially limited by computation and not bandwidth. Intel's algorithm
is more efficient in this regard, with only a 1.3-fold decrease vs. our factor
of 1.4. The additional data transfers over PCIe render the GPU algorithm
uncompetitive.

To better characterize performance, we measured the exact traffic
at each socket's memory controller. Since this information is not
available from current profilers such as VTune (which use per-core
performance counters), we have developed a small kernel-mode driver to
provide access to the model-specific performance counters in the Intel i7 uncore%
\footnote{The part of the socket not associated with a particular core.}.
Uncached writes constitute the bulk of the write combiners' memory traffic
and are therefore of particular interest. They are apparently reported as
Invalid-To-Exclusive transitions and can thus be counted as the total
number of {\em reads} minus `normal' reads \cite{intelPerfGuide}.
We find that 2041 MiB are written, which corresponds to 64~Mi items $\cdot$
8 bytes per item $\cdot$ 4 passes (slightly less because our final pass
cannot use non-temporal writes when the output position is not aligned).
Surprisingly, 2272 MiB are read. The cause of the additional 10~\% is
unknown and will be investigated in future work. However, we can provide
a conservative estimate of the bandwidth utilization. Given the pure read
and write bandwidths (38687 MB/s and 28200 MB/s) measured by
RightMark \cite{rightMark}, the minimum time required to read and write
the items' 2048 MiB is 132~ms, which is 88~\% of the total measured time.
Since this calculation does not include write-to-read turnaround \cite[p.\ 486]{memSys},
there is even less room for improvement than indicated.

\section{Conclusion}
\label{sec:conclusion}

We have introduced improvements to counting sort and a novel variant
of radix sort for integer key/value pairs. Bandwidth measurements
indicate our algorithm's throughput is within 12~\% of the theoretical
optimum for the given hardware. It outperforms the recently published
results of Intel's radix sort by a factor of 1.5 and also
outpaces a Fermi GPU when data transfer overhead is included.
These results indicate that scalar, bandwidth-sensitive sorting
algorithms still have their place on current architectures.
We believe the general software write-combining technique can provide
similar speedups for other memory-intensive applications.

\appendix

\bibliographystyle{unsrt}
\bibliography{refs}

\end{document}